# Single Color Centers Implanted in Diamond Nanostructures

Birgit J. M. Hausmann[1*], Thomas M. Babinec[1*], Jennifer T. Choy[1*], Jonathan S. Hodges[2-4], Sungkun Hong[2], Irfan Bulu[1], A. Yacoby[2], M. D. Lukin[2], Marko Lončar[1]

[1] School of Engineering and Applied Sciences, Harvard University, Cambridge, MA 02138, U.S.A.

[2] Department of Physics, Harvard University, Cambridge, MA 02138, U.S.A.

[3] Department of Nuclear Science and Engineering, Massachusetts Institute of Technology, Cambridge, MA 02139, U.S.A.

[4] Department of Electrical Engineering, Columbia University, New York, NY 10027, U.S.A.

[*] These authors contributed equally to this work

## Abstract

The development of materials processing techniques for optical diamond nanostructures containing a single color center is an important problem in quantum science and technology. In this work, we present the combination of ion implantation and top-down diamond nanofabrication in two scenarios: diamond nanopillars and diamond nanowires. The first device consists of a 'shallow' implant (~20nm) to generate Nitrogen-vacancy (NV) color centers near the top surface of the diamond crystal. Individual NV centers are then isolated mechanically by dry etching a regular array of nanopillars in the diamond surface. Photon anti-bunching measurements indicate that a high yield (>10%) of the devices contain a single NV center. The second device demonstrates 'deep' (~1µm) implantation of individual NV centers into pre-fabricated diamond nanowire. The high single photon flux of the nanowire geometry, combined with the low background fluorescence of the ultrapure diamond, allows us to sustain strong photon anti-bunching even at high pump powers.

## I. Introduction

The development of a robust and practical quantum information processing system is an important problem at the interface between materials science, photonics, and atomic physics. Quantum optical communication systems based on light-emitting defects (color centers) in diamond are particularly attractive because of their ability to generate non-classical states of light. For example, color centers based on Nitrogen[1-3], Silicon[4], Carbon[5], Nickel[6], and Chromium[7] impurities have all been demonstrated to emit single photons at room temperature. Of these, the Nitrogen-Vacancy (NV) center is most attractive at this time since it can possess additional electron and nuclear spin degrees of freedom with a long coherence time that can act as a quantum memory for long distance quantum communications[8-9], quantum computing[10-14], and nanoscale magnetometry[15-17]. Efforts to identify other outstanding color centers in diamond are on-going[18].

Practical implementations of these technologies require the integration of single color centers into diamond photonic systems whose optical performance exceeds that offered by a homogenous bulk diamond crystal. Towards this end, a variety of photonic devices have already been demonstrated that engineer the optical properties of an NV center and interface the light emitted by this atomic-scale impurity to macroscopic optical systems. Some of these devices are based on broad-band antenna[3,19] or solid immersion lens[20] effects, while others are based on narrow-band cavity enhancements[21-25]. Concurrently, there has been interest in engineering the spatial distribution of single color centers in diamond via ion implantation on a large scale. Techniques based on blanket implantation at different dosages[26-27], focused ion implantation[28], and implantation through nanoscale apertures[29-30] have all been demonstrated thus far to generate single color centers in a bulk crystal. Future scalable quantum technologies based on color centers in diamond will benefit from complete fabrication routines that combine diamond photonic device engineering with ion implantation of single centers.

In this article we describe the fabrication and characterization of two different diamond nanostructures containing single, implanted NV centers. In Section II we demonstrate the ability to generate large and regular arrays of diamond nanopillars near the top surface of the diamond crystal via a low-energy, 'shallow' implantation of Nitrogen followed by dry etching. This deterministic fabrication technique could be used to facilitate the coupling of single NV centers to proximal nanophotonic devices in the future. In section III we demonstrate the ability to perform a high-energy, 'deep' implantation of Nitrogen at a low density into pre-fabricated diamond nanowire arrays. An NV center in the nanowire acts as a high-flux source of single photons due to an antenna effect that modifies its radiation pattern[3,31]. The pure diamond host also generates a low level of background fluorescence compared to previously demonstrated nanowire devices in type Ib material, so that strong anti-bunching is observed at high pump powers where single photon count rates are maximized. In section IV we discuss some natural extensions of these fabrication routines to future diamond photonic systems.

## II. Fabrication and Characterization of Diamond Nanopillars

Figure 1 shows a cartoon of the approach we take to realize the diamond nanopillar arrays studied in this experiment. In this scheme, high quality electronic grade type IIa CVD diamond (Element 6) with low (<5ppb) background Nitrogen content was implanted with $^{15}N$ ions at an energy of 14keV and a dosage of $1.25*10^{12}$ /cm$^2$. Stopping Range of Ions in Matter (SRIM) calculations[32] project a Nitrogen layer ~20nm below the diamond surface. The sample was then annealed at 750º C for 2 hours in high vacuum (<$10^{-6}$ Torr) in order to mobilize vacancies and generate a shallow layer of NV centers. Arrays of circular shaped mask (XR electron beam resist, Dow Corning) with ~65nm radius were defined on the top surface using electron beam lithography tool (Elionix). An oxygen dry etch, whose conditions were similar to those described elsewhere[31], was applied for one minute and generated ~200nm-tall pillars on the top of the diamond surface. Finally, the sample was then placed in a hydrofluoric acid wet etch for approximately 20 seconds to remove the residual mask layer and then in a 1:1:1 mixture of Sulfuric, Nitric, and Perchloric acid at 400ºC for ~30 minutes to clean the sample. A scanning electron micrograph of a typical nanopillar array is presented in figure 2a.

The sample was then characterized with a home-built confocal microscope system as described elswhere[3]. An image of the fluorescence emitted under 532nm continuous wave (CW) excitation is presented in figure 2b. It shows a regular array of bright white spots corresponding to the devices in figure 2a. Variations in the overall intensity are observed due to the random process that embeds NV centers inside each nanopillar device. Measurements of the photoluminescence spectrum of the devices show the characteristic zero-phonon line as well as broad phonon sideband of an embedded NV center (Fig. 2c). In most devices, however, deviation in the position of the zero-phonon line (ZPL) from 637nm was observed. This appears to be consistent with reported ZPL shifts for diamond nanocrystals[33].

We then quantified the number of NV centers in individual nanopillar devices in this 10x10 array. In general, the number of quantum emitters may be obtained from measurements of the fluorescence intensity autocorrelation function $g^{(2)}(\tau) = < I(t) * I(t+\tau) > / < I(t) >^2$ using a beam-splitter, two single photon counting avalanche photodiodes (Perkin Elmer), and a time-to-amplitude converter (PicoHarp) in the Hanbury Brown and Twiss configuration[1]. This $g^{(2)}(\tau)$ function is obtained from the number of self-coincidences C.C.($\tau$) as a function of delay time $\tau$ according to $g^{(2)}(\tau) = C.C.(\tau) / (R_1 R_2 w T)$, where $R_1$ and $R_2$ are the photon count rates on channels 1 and 2, respectively, w is the bin width, and T is the integration time. A single NV center cannot emit two photons at the same time, and the number of self-coincidence counts at zero delay $g^{(2)}(0)$ is therefore expected to be zero in the absence of background sources. The observed contrast decreases according to the formula $g^{(2)}(0) = 1 – 1/N$ for additional emitters N. Ten of the seventeen devices that we tested showed strong photon anti-bunching $g^{(2)}(0) < 0.5$ (Fig. 2d), indicating that a single NV center is embedded in the nanopillar. The other seven devices were characterized by $0.5 < g^{(2)}(0) < 0.7$, which is consistent with the presence of two or three NV centers in the device. The remaining devices were untested, which suggests the need for developing automated characterization tools or perhaps wide-field microscope approaches to examine many devices in parallel. We can still conclude that at least ten of the one hundred total devices contained a single NV center, so that this fabrication routine may be implemented with high yield (>10%) operating in the single photon regime.

## III. Fabrication and Characterization of Diamond Nanowires

The results from section II demonstrate that NV centers may be implanted into diamond nanostructures fabricated with top-down techniques, and that the background fluorescence generated by processing diamond with standard nanofabrication tools is not prohibitively high for room-temperature studies of single color centers. We now modify the approach slightly for the diamond nanowire geometry, which has been shown to offer significant enhancement in the collection of single photons from an NV center compared to bulk crystals due to modification of the far-field radiation pattern[3,31]. A cartoon of the procedure used in order to realize large arrays of ~200nm diameter nanowires with ~2μm height is shown in figure 3. The fabricated devices were then implanted with $^{15}N$ at 1.7MeV and $1*10^9$ /cm$^2$ dosage and annealed at 750º C in high vacuum (<$10^{-6}$ Torr) for 2 hours. SRIM calculations indicate that this produces a layer of NV centers ~1.0μm below the diamond surface. A scanning electron micrograph of the sample is presented in figure 4.

The yield of devices containing an NV center in the diamond nanowire is relatively low due to the reduced implantation dosage. In order to identify a successfully implanted device, we therefore scanned over large (~15x15) sections of the array at high powers (~3mW) with a 532nm CW laser in order to bleach the background fluorescence from the nanowire devices. Implanted nanowires demonstrated sustained brightness due to the photo-stability of the NV center. Figure 5 shows a series of intensity auto-correlation measurements taken from one representative device at a series of increasing pump powers. Photon anti-bunching in the nanowire fluorescence as strong as $g^{(2)}(0) \sim 0.06$ is possible in the best case (Fig. 5b), which represents a ~5-fold reduction in the multi-photon probability compared to nanowire devices demonstrated in type Ib material[3]. In addition, the photon anti-bunching $g^{(2)}(0) < 0.5$ is sustained at pump powers where the single photon signal is saturated.

The light in – light out curve for this device demonstrates that an implanted nanowire may act as a high-flux single photon source (Fig. 6b). Contributions to the nanowire fluorescence come from single photons, S, from an individual NV center that leads to non-classical correlations at zero time delay $\tau = 0$, as well as background fluorescence, B, emitted with Poisson statistics that reduces the observed contrast. For a given pump power, the relative contributions of S and B to the total fluorescence are encoded in the contrast in the anti-bunching measurements according to[2] $g^{(2)}(0) = 1 - \rho^2$, where $\rho = S/(S+B)$. We therefore measured both the total number of photon counts per second from the nanowire as well as $g^{(2)}(0)$ for different pump powers P (Fig. 6b). The number of single photons collected from the nanowire was observed to turn on sharply at low pump powers and saturate at high powers according to the familiar model $S(P) = \frac{CPS_{Sat}}{1 + P_{Sat}/P}$ for the fluorescence of an NV center[1]. We observed that $CPS_{Sat}$ = 304,000 photon counts per second was the maximum possible number of photons collected and $P_{Sat}$ = 0.34mW was the pump power scale on which the NV center emission was saturated. This is consistent with the efficient excitation and extraction of single photons from an NV center that was previously demonstrated in type Ib nanowire antennas in our confocal microscope system[3].

Additional information about the optical properties of an NV center in a diamond nanowire may be obtained by triggering the emission of single photons via pulsed excitation of the NV center. We have constructed a versatile system for this task[34]. Ultrafast (~200fs) pulses generated by a Ti:Sapphire (Coherent Mira 800-F) laser were used to generate supercontinuum white light using a photonic crystal fiber (Newport, SCG-800). A pump wavelength of ~800nm maximized the overall spectral density at green wavelengths, which was then isolated using band-pass filters in the range 510-540nm (Semrock). Since the fundamental repetition rate of the Ti:Sapphire pulse train (76MHz) is comparable to the radiative rate of an NV center in a diamond nanowire (~60-80MHz)[3], we reduced the repetition rate of the pump to ~10.8 MHz using an electro-optic modulator (ConOptics, Model 350) prior to launching into the photonic crystal fiber.

This pulsed excitation scheme was then used to observe the full temporal dynamics of the fluorescence from the diamond nanowire antenna. Pulsed intensity auto-correlation measurements show a series of spikes in coincidence counts at times separated by an integer number of laser repetition cycles (Fig. 7a). Strong suppression of the central peak $g^{(2)}(0) \sim 0.16$ is again observed at zero delay, which is consistent with the extent of the photon anti-bunching observed using CW excitation. We were also able to measure the fluorescence lifetime of the NV center directly from the decay in the nanowire fluorescence after pulsed excitation (Fig. 7b). An excellent fit is obtained from a bi-exponential function using a fast time constant $\tau_{bg} \sim 1.4 \pm 0.1$ ns, which corresponds to the decay of the background fluorescence, and a slow time constant $\tau_{nv} \sim 13.7 \pm 0.2$ ns, which corresponds to the fluorescence decay of the NV center in the nanowire. This value of $\tau_{nv}$ is consistent with previously reported values inferred from the width of the anti-bunching dip in CW studies of type Ib nanowires[3].

## V. Conclusions and Future Directions

We have demonstrated two novel materials processing techniques to implant color centers in diamond nanostructures, and both of these could be an important part of future quantum photonic systems based on diamond. For example, embedding the nanopillar arrays presented in Section II in a metal layer could allow for plasmon-enhanced single photon emission[35]. Device arrays such as these could also offer convenient, evanescent coupling to other proposed photonic crystal cavities in semiconductor material systems for cavity quantum electrodynamics studies[25,36-37]. In either case, the observed scalability provided by this system is an attractive resource for the development of more complex and integrated device architectures.

There are also several natural extensions of the deep implantation of color centers into the nanowires shown in Section III. Here, the outstanding combination of high directionality of emission from the nanowire antenna combined with low background fluorescence in the pure diamond crystal could allow us to greatly reduce the requirements on the optical systems used to probe a single color center. Indeed, preliminary work in one our systems indicate that it is possible to observe anti-bunching as strong as $g^{(2)}(0) \sim 0.1$ in a confocal microscope with a lower numerical aperture of NA ~ 0.6, though at slightly reduced collection efficiency. We should be able to go one step further and integrate classical lightwave technology with a quantum optical light source by coupling the emission of a single NV center directly to a lensed optical fiber (NA ~ 0.4) via the diamond nanowire antenna. Finally, the nanowire architecture provides a general setting for conducting studies of the low-temperature properties of an NV center (stability of optical transitions, effects of strain, etc…) in diamond nanophotonic structures.

## Acknowledgements

We thank Patrick Maletinsky and Jero Maze for assistance with annealing the diamond used in this experiment, Fedor Jelezko and Helmut Fedder for helpful discussions about the implementation of the pulsed excitation system, and Daniel Twitchen from Element Six for helpful discussions and for diamond samples. T. M. B. acknowledges support from the NDSEG and NSF fellowships, and J. T. C. acknowledges support from the NSF graduate research fellowship. Devices were fabricated in the Center for Nanoscale Systems (CNS) at Harvard. This work was supported in part by Harvard's Nanoscale Science and Engineering Center (NSEC), NSF NIRT grant (ECCS-0708905), and by the DARPA QuEST program.

## References

1 Kurtsiefer, C., Mayer, S., Zarda, P. & Weinfurter, H. Stable solid-state source of single photons. *Phys. Rev. Lett.* **85**, 290-293 (2000).

2 Beveratos, A., Brouri, R., Gacoin, T., Poizat, J.-P. & Grangier, P. Nonclassical radiation from diamond nanocrystals. *Phys. Rev. A* **64**, 061802(R) (2001).

3 Babinec, T. M., Hausmann, B. J. M., Khan, M., Zhang, Y., Maze, J. R., Hemmer, P. R. & Loncar, M. A diamond nanowire single-photon source. *Nature Nano.* **5**, 195-199 (2010).

4 Wang, C., Kurtsiefer, C., Weinfurter, H. & Burchard, B. Single photon emission from SiV centres in diamond produced by ion implantation. *J. Phys. B: At. Mol. Opt. Phys.* **39**, 37-41 (2006).

5 Naydenov, B., Kolesov, R., Batalov, A., Meijer, J., Pezzagna, S., Rogalla, D., Jelezko, F. & Wrachtrup, J. Engineering single photon emitters by ion implantation in diamond. *Appl. Phys. Lett.* **95**, 181109 (2009).

6 Gaebel, T., Popa, I., Gruber, A., Domhan, M., Jelezko, F. & Wrachtrup, J. Stable single-photon source in the near infrared. *New J. Phys.* **6**, 98 (2004).

7 Aharonovich, I., Castelletto, S., Simpson, D. A., Stacey, A., McCallum, J., Greentree, A. D. & Prawer, S. Two-level ultrabright single photon emission from diamond nanocrystals. *Nano Lett.* **9**, 3191-3195 (2009).

8 Childress, L., Taylor, J. M., Sorensen, A. S. & Lukin, M. D. Fault-tolerant quantum communicatin based on solid-state photon emitters. *Phys. Rev. Lett.* **96**, 070504 (2006).

9 Childress, L., Taylor, J. M., Sorensen, A. S. & Lukin, M. D. Fault-tolerant quantum repeaters with minimal physical resources and implementations based on single-photon emitters. *Physical Review A* **72**, 052330 (2005).

10 Jelezko, F., Gaebel, T., Popa, I., Gruber, A. & Wrachtrup, J. Observation of coherent oscillations in a single electron spin. *Phys. Rev. Lett.* **92**, 076401 (2004).

11 Jelezko, F., Gaebel, T., Popa, I., Domhan, M., Gruber, A. & Wrachtrup, J. Observation of coherent oscillation of a single nuclear spin and realization of a two-qubit conditional quantum gate. *Phys. Rev. Lett.* **93**, 130501 (2004).

12 Gurudev Dutt, M. V., Childress, L., Jiang, L., Togan, E., Maze, J., Jelezko, F., Zibrov, A. S., Hemmer, P. R. & Lukin, M. D. Quantum register based on individual electronic and nuclear spin qubits in diamond. *Science* **316**, 1312-1316 (2007).

13 Neumann, P. *et al.* Quantum register based on coupled electron spins in a room-temperature solid. *Nature Physics* **6**, 249-253 (2010).

14 Togan, E. *et al.* Quantum entanglement between an optical photon and a solid-state qubit. *Nature* **466**, 730-735 (2010).

15 Maze, J. R. *et al.* Nanoscale magnetic sensing with an individual electronic spin in diamond. *Nature* **455**, 644-648 (2008).

16 Balasubramanian, G. *et al.* Nanoscale imaging magnetometry with diamond spins under ambient conditions. *Nature* **455**, 648-652 (2008).

17 Taylor, J. M., Capellaro, P., Childress, L., Jiang, L., Budker, D., Hemmer, P. R., Yacoby, A., Walsworth, R. & Lukin, M. D. High-sensitivity diamond magnetometer with nanoscale resolution. *Nature Phys.* **4**, 810-816 (2008).

18 Weber, J. R., Koehl, W. F., Varley, J. B., Janotti, A., Buckley, B. B., Walle, C. G. V. d. & Awschalom, D. D. Quantum Computing with Defects. *PNAS* **107**, 8513-8518 (2010).

19 Schietinger, S., Barth, M., Aichele, T. & Benson, O. Plasmon-enhanced single photon emission from a nanoassembled metal/diamond hybrid structure at room temperature. *Nano Letters* **9**, 1694-1698 (2009).

20 Hadden, J. P., Harrison, J. P., Stanley-Clarke, A. C., Marseglia, L., Ho, Y.-L. D., Patton, B. R., O'Brien, J. L. & Rarity, J. G. Strongly enhanced photon collection from diamond defect centres under micro-fabricated integrated solid immersion lenses. *arXiv:1006.2093v2* (2010).

21 Park, Y.-S., Cook, A. K. & Wang, H. Cavity QED with diamond nanocrystals and silica microspheres. *Nano Lett.* **6**, 2075-2079 (2006).

22 Larsson, M., Dinyari, K. N. & Wang, H. Composite optical microcavity of diamond nanopillar and silica microsphere. *Nano Lett.* **9**, 1447-1450 (2009).

23 Fu, K.-M. C., Santori, C., Barclay, P. E., Aharonovich, I., Prawer, S., Meyer, N., Holm, A. M. & Beausoleil, R. G. Coupling of nitrogen-vacancy centers in diamond to a GaP waveguide. *Appl. Phys. Lett.* **93**, 234107 (2008).

24 Barclay, P. E., Fu, K.-M. C., Santori, C. & Beausoleil, R. G. Chip-based microcavities coupled to nitrogen-vacancy centers in single crystal diamond. *Appl. Phys. Lett.* **95**, 191115 (2009).

25 Englund, D., Shields, B., Rivoire, K., Hatami, R., Vuckovic, J., Park, H. & Lukin, M. D. Determinstic coupling of a single nitrogen vacancy center to a photonic crystal nanocavity. *arXiv:1005.2204* (2010).

26 Gruber, A., Drabenstedt, A., Tietz, C., Fleury, L., Wrachtrup, J. & Borczyskowski, C. Scanning confocal optical microscopy and magnetic resonance on single defect centers. *Science* **276**, 2012-2014 (1997).

27 Rabeau, J. R. *et al.* Implantation of labelled single nitrogen vacancy centers in diamond using 15N. *Appl. Phys. Lett.* **88**, 023113 (2006).

28 Meijer, J., Burchard, B., Domhan, M., Wittmann, C., Gaebel, T., Popa, I., Jelezko, F. & Wrachtrup, J. Generation of single color centers by focused nitrogen implantation. *Appl. Phys. Lett.* **87**, 261909 (2005).

29 Meijer, J. *et al.* Towards the implanting of ion and positioning of nanoparticles with nm spatial resolution. *Appl. Phys. A* **91**, 567-571 (2008).

30 Toyli, D. M., Weis, C. D., Fuchs, G. D., Schenkel, T. & Awschalom, D. D. Chip-Scale Nanofabrication of Single Spins and Spin Arrays in Diamond. *Nano. Lett.* **10**, 3168-3172 (2010).

31 Hausmann, B., Khan, M., Zhang, Y., Babinec, T., Martinick, K., McCutcheon, M., Hemmer, P. & Loncar, M. Fabrication of diamond nanowires for quantum information processing applications. *Diamond & Related Materials* **19**, 621-629 (2010).

32 Ziegler, J. F., Ziegler, M. D. & Biersack, J. P. *SRIM - The stopping range of ions in matter*, <www.srim.org> (2008).

33 Shen, Y., Sweeney, T. M. & Wang, H. Zero-Phonon Linewidth of Single Nitrogen Vacancy Centers in Diamond Nanocrystals. *Phys. Rev. B* **77**, 033201 (2008).

34 Jelezko, F. *Private communication* (2010).

35 Hausmann, B. J. M., Bulu, I., Babinec, T. M., Choy, J., Khan, M., Hemmer, P. & Loncar, M. in *CLEO/QELS* (2010).

36 Barclay, P. E., Fu, K.-M., Santori, C. & Beausoleil, R. G. Hybrid photonic crystal cavity and waveguide for coupling to diamond NV centers. *Opt. Express* **17**, 9588-9601 (2009).

37 McCutcheon, M. W. & Lončar, M. Design of a silicon nitride photonic crystal nanocavity with a quality factor of one million for coupling to a diamond nanocrystal. *Optics Express* **16**, 19136-19145 (2008).

## Figure Captions

**Figure 1.** Approach to isolate implanted NV centers in large arrays of diamond nanopillars fabricated in high-purity single crystal diamond substrate: **(a)** start with clean type IIa diamond, **(b)** low-energy Nitrogen ion implantation plus annealing activates a thin surface layer of NV centers, **(c)** electron beam lithography resist is deposited on the diamond surface, **(d)** electron beam lithography defines an etch mask, **(e)** reactive ion etching mechanically isolates individual NV centers in the device, and finally **(f)** HF wet etch of the diamond reveals the structures.

**Figure 2. (a)** SEM image of the array of nanopillars of height ~250nm and radius ~65nm. Inset shows the profile of an individual particle, with scale bar 200nm. **(b)** Confocal microscope image of a 10x10 array of diamond nanopillars. The scale bar is equal to the 2μm particle spacing. **(c)** Photoluminescence spectrum of the nanopillar highlighted in part (b). The zero-phonon line of this device corresponds to ~634nm. **(d)** Intensity autocorrelation function of the fluorescence emitted from the device highlighted in part (b) shows strong photon anti-bunching $g^{(2)}(0) \sim 0.08$ due to the presence of a single embedded NV center.

**Figure 3.** Approach to implant individual NV centers in large arrays of diamond nanowire antennas made from high-purity single crystal diamond: **(a)** start with clean type IIa diamond, **(b)** electron beam lithography resist is deposited on the diamond surface, **(c)** electron beam lithography defines an etch mask, **(d)** HF wet etch of the diamond reveals the structures, **(e)** high-energy ion implantation of Nitrogen, and finally **(f)** annealing activates NV centers.

**Figure 4. (a)** SEM image of an array of nanowires of height ~2μm and diameter ~200nm examined in this experiment. Thousands of devices can be fabricated in parallel due to high-throughput processing techniques used in this experiment. **(b-c)** These insets show the profile of individual nanowires in this large array. The scale bar in these images is 200nm.

**Figure 5.** Intensity autocorrelation data for a single NV center implanted in a diamond nanowire shows sustained anti-bunching at increasing excitation powers due to high single photon collection efficiency and low background: **(a)** 50μW, **(b)** 100μW, **(c)** 250μW, **(d)** 1000μW. The number on each plot correspond to the value $g^{(2)}(0)$ after normalizing the coincidence count data and without background subtraction. Pump power is measured before the microscope objective.

**Figure 6. (a)** Data points correspond to the values of the normalized $g^{(2)}(0)$ data presented in figure 5. Dashed curve is the expected level of coincidence counts $1 - \rho^2$ based on the measured saturation data presented in part (b), where $\rho = S / (S+B)$, S is NV center single photon count rate, and B is background fluorescence count rate. **(b)** Total nanowire fluorescence from the same device is presented in red, the background fluorescence obtained from contrast in photon anti-bunching data in part (a) is shown in blue, and the remainder of the net single photon counts from an embedded NV center is shown in black. A fit of the NV center fluorescence to the saturation model presented in section III is given by the dashed black line. Pump power is measured before the microscope objective.

**Figure 7. (a)** Photon anti-bunching measurement of a diamond nanowire using pulsed excitation (65μW average pump power). Strong suppression of the central peak at zero delay $g^{(2)}(0) \sim 0.16$ is observed. **(b)** Measurement of the fluorescence lifetime via fluorescence decay is shown in black (14μW average pump power). The red line corresponds to a bi-exponential fit to the fluorescence decay with time constants $\tau_{bg} = 1.5$ ns for backgounrd fluorescence and $\tau_{nv} = 13.7$ ns for the NV center fluorescence.

**Figure 1**

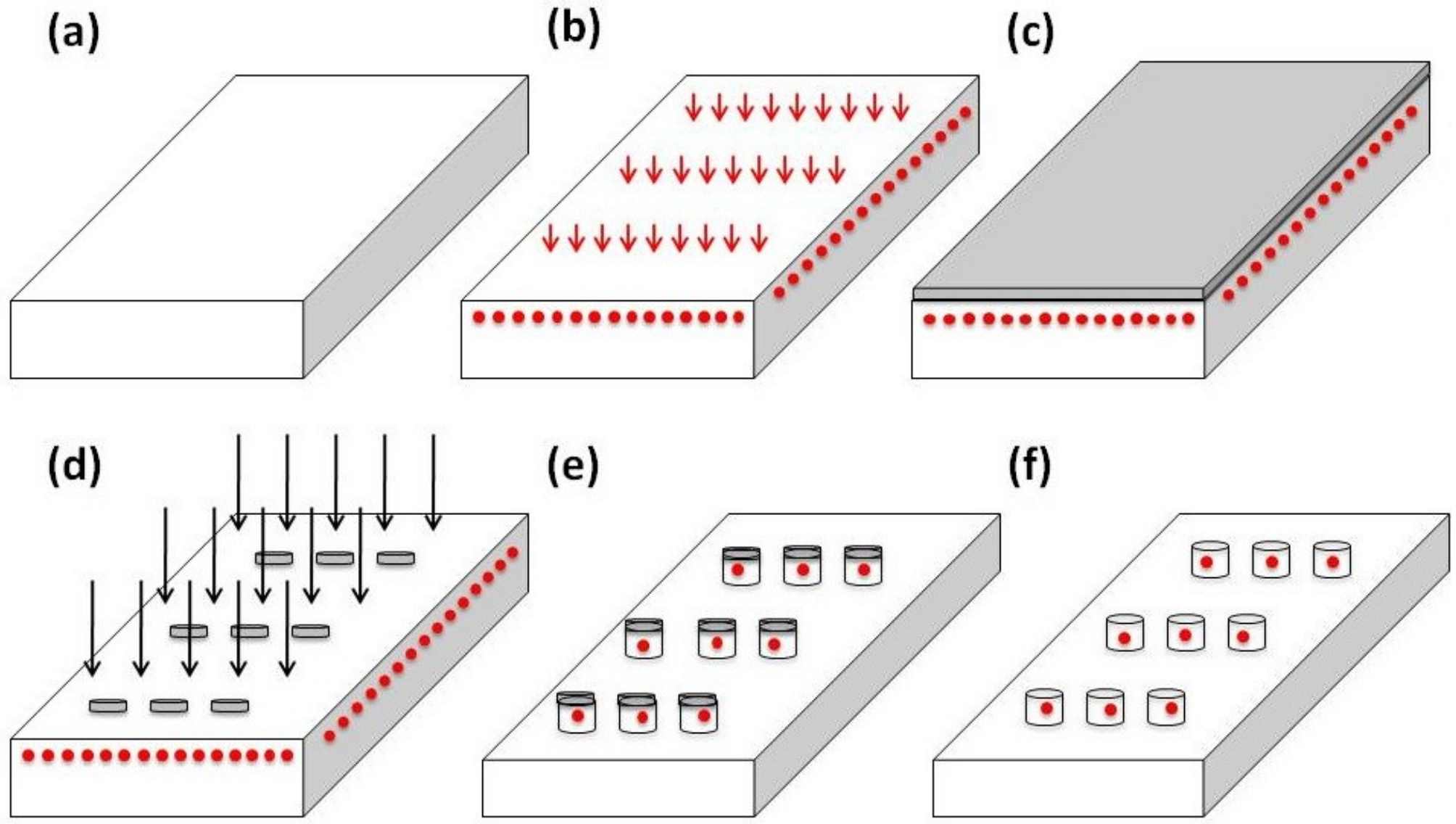

**Figure 2**

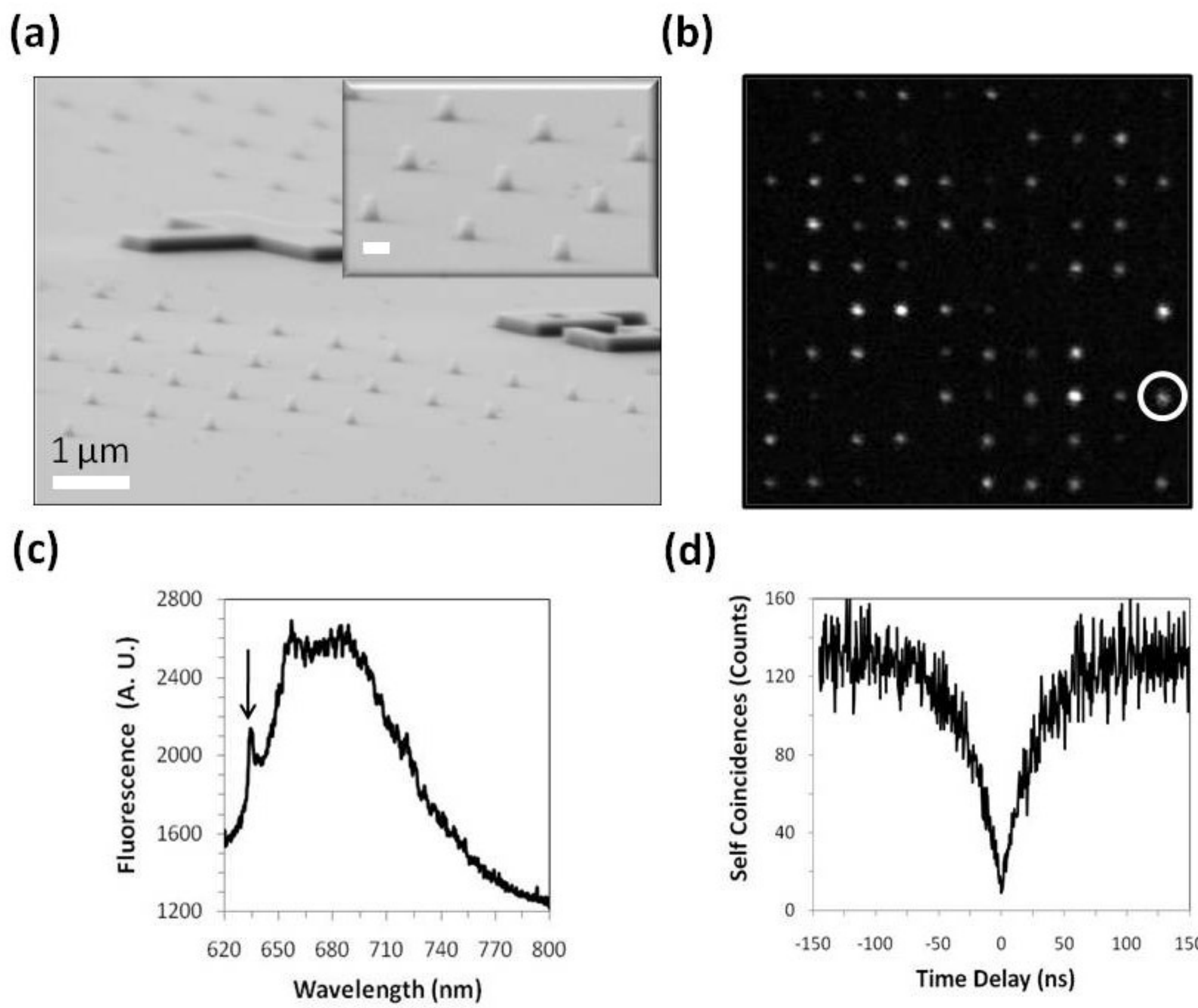

**Figure 3**

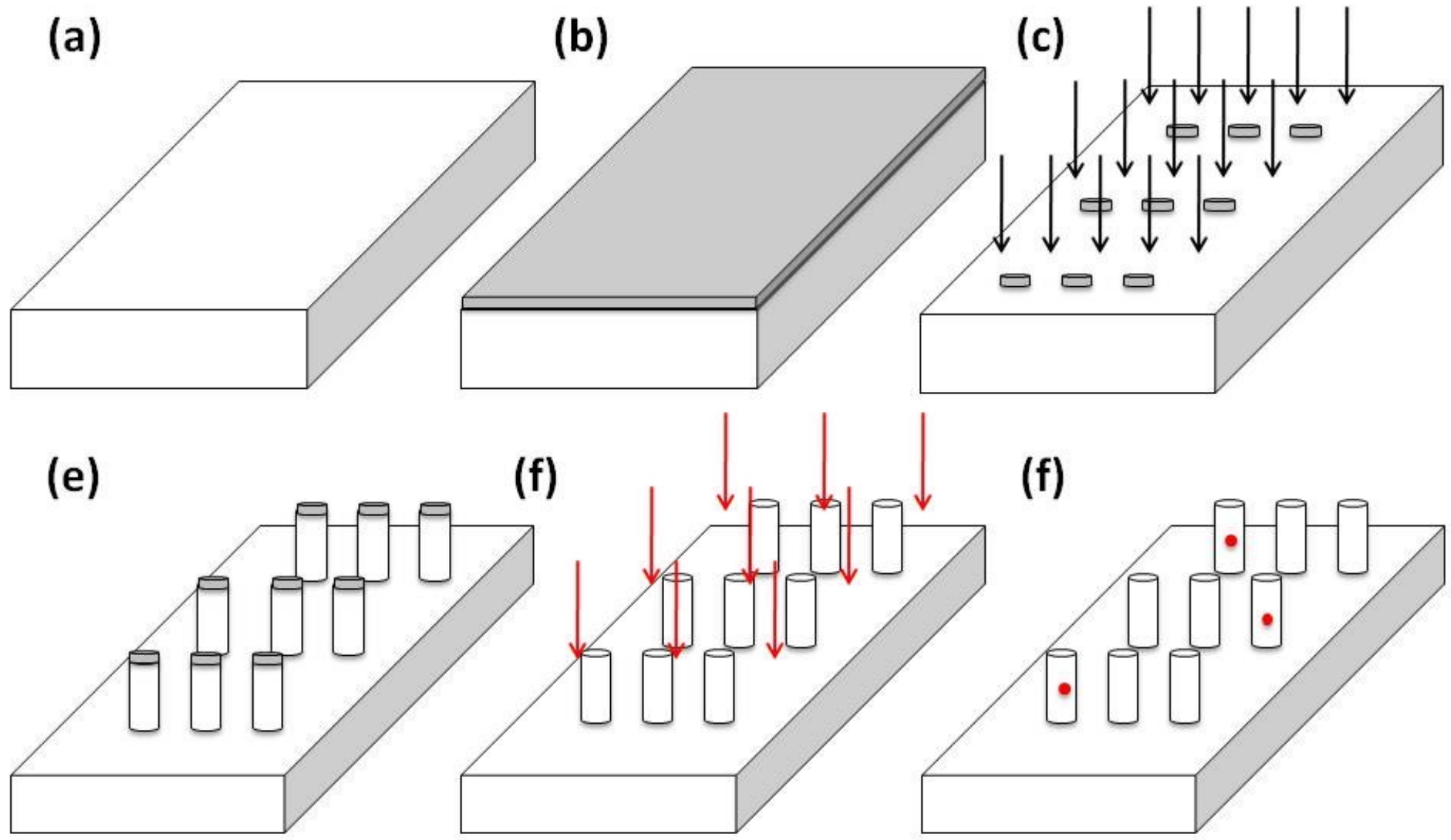

**Figure 4**

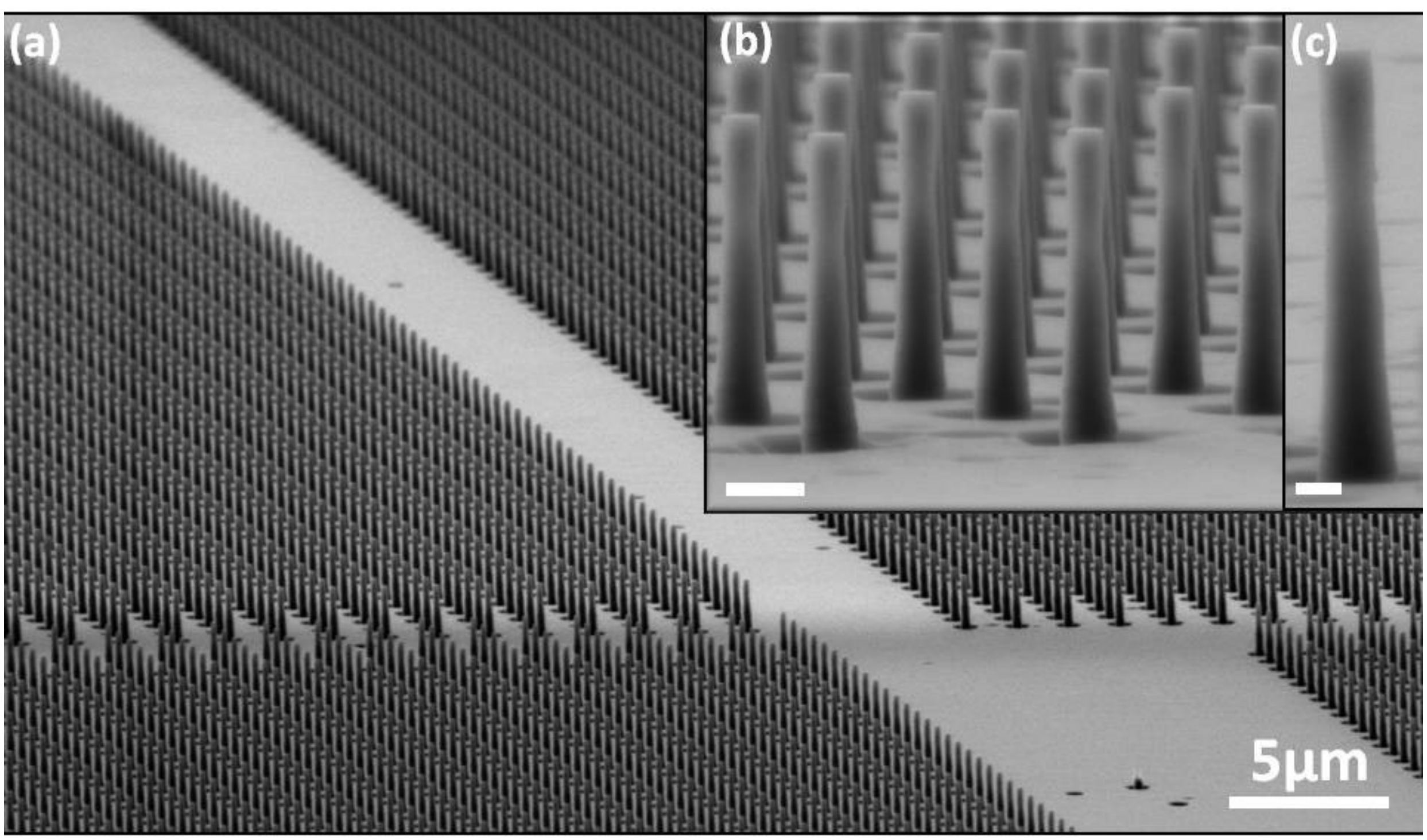

**Figure 5**

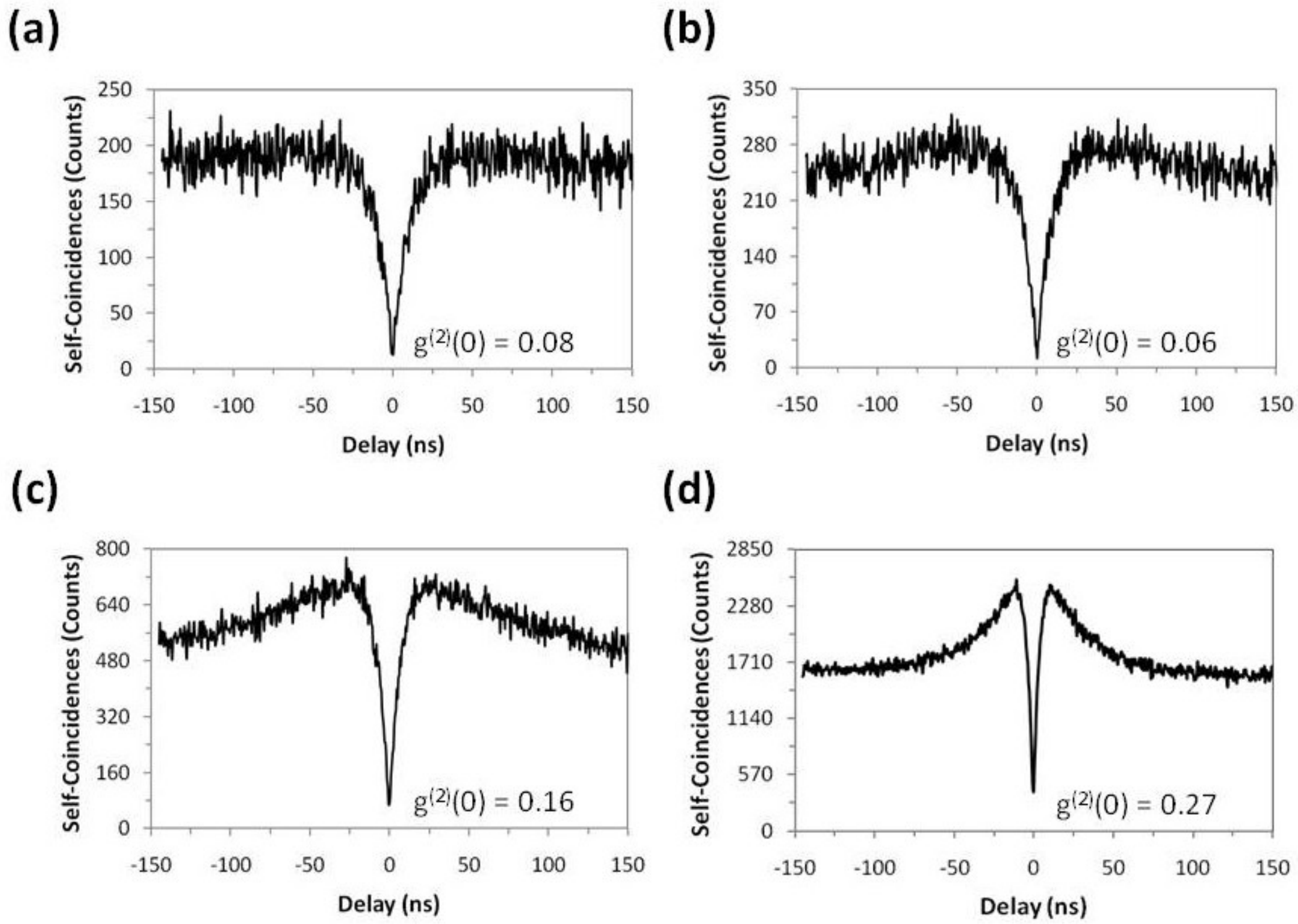

**Figure 6**

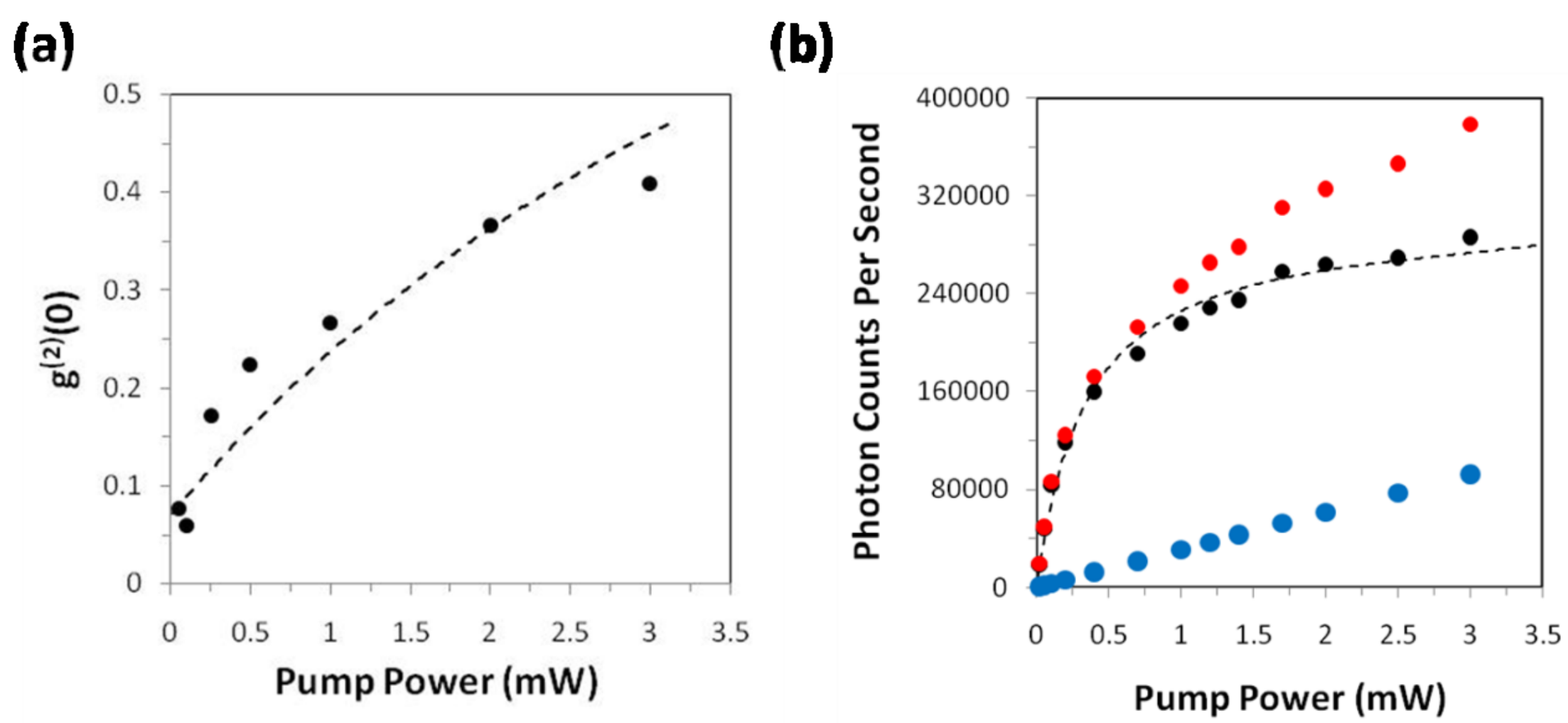

**Figure 7**

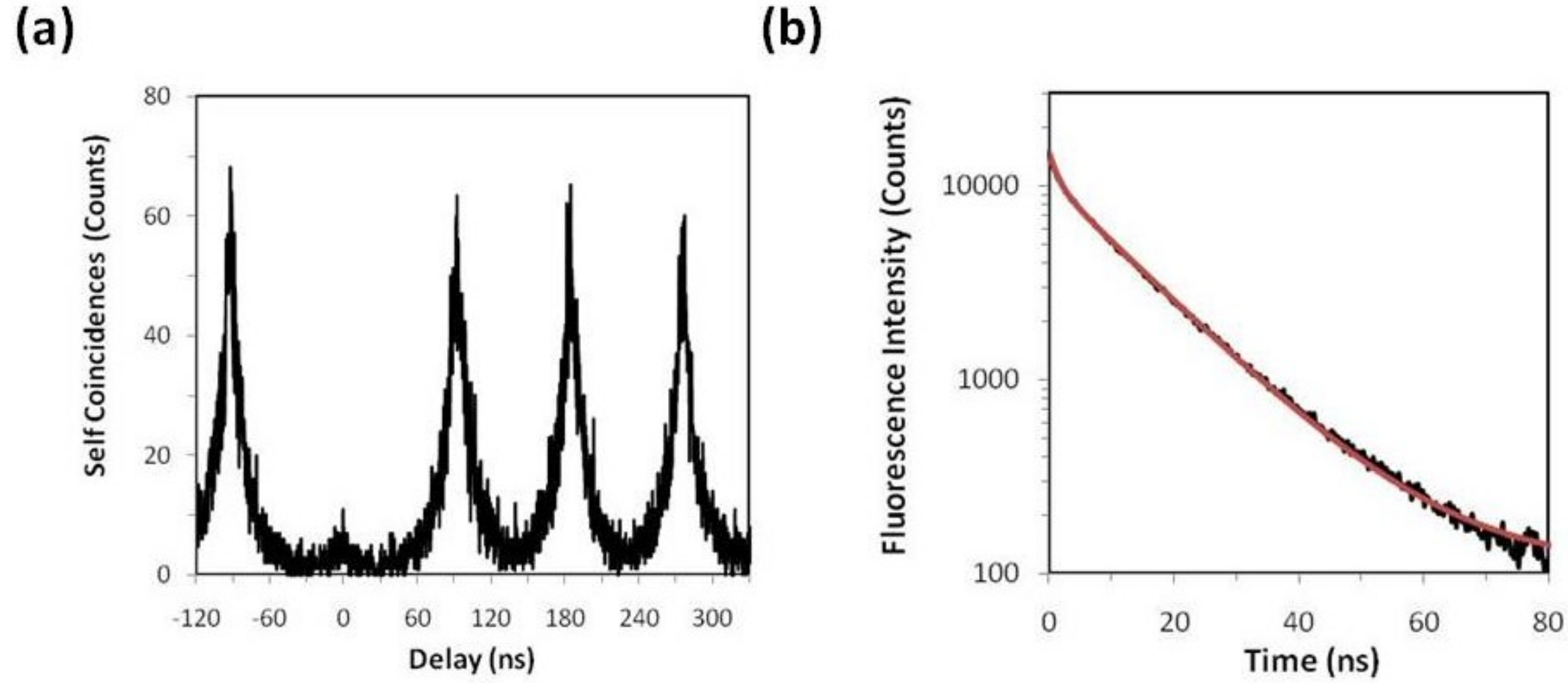

(a)
Self Coincidences (Counts)
80
60
40
20
0
-120
-60
0
60
120
180
240
300
Delay (ns)
(b)
Fluorescence Intensity (Counts)
10000
1000
100
0
20
40
60
80
Time (ns)